\documentclass[floatfix,twocolumn,aps,prd,showpacs,amsmath,amssymb]{revtex4}
\usepackage{epsfig}
\usepackage[latin1]{inputenc}

\usepackage[dvips]{color}

\begin{document}

\title{ Unitarity of theories containing  fractional powers of the d'Alembertian operator}

\author{E. C. Marino$^1$, Leandro O. Nascimento$^{1,3}$, Van S\'ergio Alves$^{1,2}$, and C. Morais Smith$^3$}
\affiliation{$^1$Instituto de F\'\i sica, Universidade Federal do Rio de Janeiro, C.P.68528, Rio de Janeiro RJ, 21941-972, Brazil \\
$^2$Faculdade de F\'\i sica, Universidade Federal do Par\'a, Av.~Augusto Correa 01, 66075-110, Bel\'em, Par\'a, Brazil \\
$^3$Institute for Theoretical Physics, Utrecht University, Leuvenlaan 4, 3584CE Utrecht, The Netherlands}

\date{\today}

\begin{abstract}
 
We examine the unitarity of a class of generalized Maxwell U(1) gauge theories in (2+1) D containing the
pseudodifferential operator $\Box^{1-\alpha}$, for $\alpha \in [0,1)$. We show that only Quantum Electrodynamics 
(QED$_3$) and its generalization
known as Pseudo Quantum Electrodynamics (PQED), for which $\alpha  =0$ and $\alpha = 1/2$, respectively, satisfy unitarity. The latter plays
an important role in the description of the electromagnetic interactions of charged particles confined to a
plane, such as in graphene or in hetero-junctions displaying the quantum Hall effect. 
\end{abstract}

\pacs{11.15.-q, 11.10.Lm, 11.55.Bq}

\maketitle

\section{Introduction}

Unitarity is an important necessary condition for the consistency of any quantum theory.
Consider the time evolution operator $U(t,0)$, defined as
\begin{equation}
|\Psi(t)\rangle = U(t,0) |\Psi(0)\rangle,
\end{equation}
where $|\Psi(t)\rangle$ is the state-vector at instant  $t$.
The unitarity of the time-evolution operator,
namely the property $U^{\dagger}U = U U^{\dagger} = I$, where $I$ is the identity operator,
guarantees that the norm of the state-vectors, chosen equal to one, is preserved in time. Since the
state-vector can be expanded in the eigenstates of any observable $A$, it follows that its norm  is equal to
the sum of the probabilities 
for the possible outcomes of any measurement of $A$. Unitarity implies that this sum of probabilities 
remains equal to one at any time, an essential condition for the probabilistic description of a system.
For a time-independent Hamiltonian, we have $U(t,0) = \exp(-iHt)$. Unitarity
then  implies that the Hamiltonian is a hermitian operator and therefore the energy eigenvalues are real.
This property and the conservation of the sum of probabilities are crucial conditions for the stabillity of
a quantum-mechanical system \cite{peskin}.

Another consequence of the unitarity of the time-evolution operator is that the scattering matrix, which
connects the asymptotic states after a scattering event to the ones before it, must also be unitary.
Assuming the completeness of the asymptotic states, then it follows that the $S$-matrix  elements form
a matrix representation of a unitary scattering operator  $S=1+i\, T$.
Unitarity of the $S$-operator, namely, $S^{\dagger}S=1$, implies 
\begin{equation}
i \,(T^{\dagger}-T)=T^{\dagger}T.
\label{0}
\end{equation}
This relation leads to  the optical theorem, which relates the forward scattering amplitude to the total cross section of the scatterer. A very convenient way of testing the
consistency of a theory is then provided by the optical theorem, which is satisfied by unitary theories (for a nice review about the optical theorem see the Ref.~\cite{matthew}).

In this paper, we examine the unitarity of a class of generalized Maxwell U(1) gauge theories in (2+1)D by using the optical theorem. For an appropriate choice of the gauge, the equations of motion for these theories are $\Box^{1-\alpha}A_{\mu}=0$, for any $\alpha \in [0,1)$. We show that only the choices $\alpha=0$ or $\alpha=1/2$ corresponding, respectively, to QED$_3$ and the so-called pseudoQED (PQED)  provide a self-consistent solution to the optical theorem. Particularly, the choice $\alpha=1/2$ is also consistent with the Huygens principle Ref.~\cite{huygens}. The unitarity of PQED is first proven at the tree level, and then for the interacting case.

The outline of this paper is the following: In Sec.~II we revise the PQED and propose its generalization to any $\alpha$. In Sec.~III we show that only $\alpha=0$ or $\alpha=1/2$ are possible choices in order to obtain a self-consistent solution of the optical theorem. Both cases are considered at the tree level, with no source term in the equation of motion. In Sec.~IV we use the RPA approach to show that the version of PQED used to describe the electronic interaction in graphene is also unitary. In Sec.~V we adopt perturbation theory up two loop to show that the PQED is an unitary theory. 

\section{The PQED and its Generalizations}

\subsection{The Derivation of PQED}

The discovery of condensed matter systems with physical properties that are essentially two-dimensional has
fostered the investigation of (2+1)D theories, which could appropriately describe them. Among these we
find the $\rm{GaAs}$  quantum wells exhibiting the quantum Hall effect, the high-Tc cuprates and graphene \cite{review}. In such
systems, a crucial issue is the description of the electronic interaction, which naturally
is electromagnetic (EM). For this matter, one must consider that
the interaction among the electrons is usually mediated by a (spacially) three-dimensional field in spite of the fact that
the electron kinematics is confined to a plane. For the sake of convenience, simplicity and elegance, however,
it is preferable to provide a completely  (2+1)-dimensional description of the real electromagnetic interaction
among the electrons. This is achieved \cite{Kovner,marinosup1,marinorubenssup,Teber} by
a theory, coined Pseudo Quantum Electrodynamics (PQED), which was also used in the bosonization of the massless
Dirac field in (2+1)D \cite{marinosup2}. Dynamical mass generation for massless electrons also was studied for this model \cite{VLWJF}.  

In this section, for the sake of completeness, we review the main steps of the derivation contained in Ref.~\cite{marinosup1}.
We start from standard QED$_4$, in (3+1)D:
\begin{eqnarray}
{\cal L}_{QED}= -\frac{1}{4} F_{\mu \nu} F^{\mu\nu} -e\,j^{\mu}_{3+1}\, A_{\mu}
+ {\cal L}_m,
\label{1a}
\end{eqnarray}
where $j^{\mu}_{3+1}$ and $ {\cal L}_m$ are, respectively, the electronic  current and kinetic Lagrangian. $A_{\mu}$ is the gauge field, $F^{\mu\nu}$ is the usual field-strength tensor.

The electromagnetic field induces an effective current-current interaction on the electrons, which is captured
by the functional (in Euclidean space)
\begin{eqnarray}
Z_{QED}[j^{\mu}_{3+1}]= Z_0^{-1}  \int D A_\mu \exp\left \{-\int d^4\xi {\cal L}_{QED} \right \},
\label{2a}
\end{eqnarray}
where $\xi=(x,y,z,\tau)$ and $Z_0$ is a normalization constant which guarantees that $Z[0]=1$.
The functional integration above can be carried out by including a gauge fixing term, yielding
\begin{eqnarray}
Z_{QED}[j^{\mu}_{3+1}]= \exp\left \{-\frac{e^2}{2}\int d^4\xi d^4\xi' \ j^{\mu}_{3+1}(\xi) \right .
\nonumber\\ \times \left .
 G_{QED}^{\mu\nu}(\xi-\xi')j^{\nu}_{3+1}(\xi')\right \},
\label{3a}
\end{eqnarray}
where $G_{QED}^{\mu\nu}$ is the Euclidean propagator of the electromagnetic field,
which is given by
\begin{eqnarray}
 G_{QED}^{\mu\nu}(\xi-\xi') = \delta^{\mu\nu} \int \frac{d^4 k}{(2\pi)^4}\frac{ e^{i k \cdot(\xi-\xi')}}{  k^2} + {\rm gt},
\label{4a}
\end{eqnarray}
where ${\rm gt}$ stands for ``gauge dependent terms''. These, by the way, do not contribute for Eq.~(\ref{3a}).

We now introduce the fact that the electrons are supposed to move on a plane at $z=0$, thus forming a spatially
two-dimensional system.  The electronic current, accordingly, is given by
\begin{eqnarray}
j^{\mu}_{3+1}(\xi) = \left \{ \begin{array}{ll}
                                              j^\mu (x,y, \tau) \delta(z), & \mu = 0, 1, 2,\\
                                                           0,                                      & \mu = 3.
                                      \end{array} \right . 
\label{5a}
\end{eqnarray}
Inserting Eq.~(\ref{5a}) in Eq.~(\ref{3a}) and integrating over $z$ and $z'$, we get
\begin{eqnarray}
Z_{QED}[j^{\mu}]= \exp\left \{-\int d^3\eta d^3\eta'\ j^{\mu}(\eta) \right .
\nonumber\\ \times \left .
 G_{QED}^{\mu\nu}(\eta-\eta'; z = z'=0)j^{\nu}(\eta')\right \},
\label{6a}
\end{eqnarray}
where $\eta=(x,y,\tau)$ and
\begin{eqnarray}
 G_{QED}^{\mu\nu}(\eta-\eta'; z = z'=0) =
\frac{ \delta^{\mu\nu}}{8 \pi^2 |\eta-\eta'|^2} + {\rm gt}.
\label{7a}
\end{eqnarray}
The expression above is the 4-dimensional QED Euclidean propagator, calculated at $z = z'=0$.

Now comes a key step in our derivation. This is the realization that Eq.~(\ref{7a}) can be written as
a 3-dimensional Fourier  integral, namely
\begin{eqnarray}
 \frac{1}{8 \pi^2 |\eta-\eta'|^2} = \int \frac{d^3 k_{3D}}{(2\pi)^3}
\frac{ e^{i k_{3D} \cdot (\eta-\eta')}}{4 \sqrt{k^2_{3D}}}
\label{8a},
\end{eqnarray}
and this is the euclidean propagator of PQED \cite{marinosup1}, corresponding to the
strictly (2+1)-dimensional Lagrangian
\begin{eqnarray}
{\cal L}_{PQED}= -\frac{1}{4} F_{\mu \nu}\left[\frac{4}{(-\Box)^{1/2}}\right] F^{\mu\nu} -e\,j^{\mu}\, A_{\mu}
+ {\cal L}_m,
\label{9a}
\end{eqnarray}
Inserting   Eq.~(\ref{7a}) and Eq.~(\ref{8a}) in  Eq.~(\ref{6a}), we can immediately realize that
\begin{eqnarray}
Z_{QED}[j^{\mu}]= Z_0^{-1}  \int D A_\mu \exp\left \{-\int d^3\eta {\cal L}_{PQED} \right \}.
\label{10a}
\end{eqnarray}

The above derivation shows that all the electronic properties determined by QED$_4$, when projected on a
plane are described by a strictly (2+1)-dimensional  theory, namely PQED. In connection to this point,
one could argue whether PQED provides a description of the correlation functions of QED$_4$. The $A_\mu$
correlators are generated by coupling an external source $ J^{\mu}_{3+1}$ in Eq.~(\ref{2a}), namely
\[
j^{\mu}_{3+1}\rightarrow j^{\mu}_{3+1} + J^{\mu}_{3+1},
\]
and subsequently taking functional derivatives of $Z_{QED}$ with respect to this source. Assuming
it has the same structure as the electronic current given by  Eq.~(\ref{5a}), it follows that functional derivatives
with respect to the (2+1)-dimensional external source taken in PQED  will generate the projected
correlators, as it occurred with the two-point function in Eq.~(\ref{7a}).

\subsection{Generalized PQED} 

We will consider here a class of theories in  (2+1)D, which contain PQED and QED$_3$ as particular cases.
These are given by 
\begin{eqnarray}
{\cal L}= -\frac{1}{4} F_{\mu \nu}\left[\frac{4}{(-\Box)^\alpha}\right] F^{\mu\nu} -e\,j^{\mu}\, A_{\mu}
+ {\cal L}_m,
\label{1}
\end{eqnarray}
where $ 0 \leq \alpha < 1$. For
a proper choice of the gauge condition, the U(1) vector field satisfies the equation
\begin{eqnarray}
 \Box^{1-\alpha}A^{\mu} = e\,j^{\mu},
\label{2}
\end{eqnarray}
which is pseudodifferential for $\alpha \neq 0$. For $\alpha =0$, the theory above is just Maxwell
QED$_3$. 
In the previous section, we have shown that the case
$\alpha = 1/2$, namely PQED, is relevant for the description of the electromagnetic interactions of two-dimensional
systems. In this case,  Eq.~(\ref{1}) provides a full description of the real electromagnetic interaction
for electrons confined on a plane \cite{marinosup1}.

In the above Lagrangian, the first term reads
\begin{equation}
F_{\mu \nu}(\eta)\int d^3\eta' \int \frac{d^3k}{(2\pi)^3} 
\frac{e^{-i k\centerdot (\eta-\eta')}}{(k^2)^\alpha}F^{\mu \nu}(\eta'),
\label{m11}
\end{equation}
where  $k =(\textbf{k},\omega)$ (we excluded the index ``3D'' for simplicity) and $\eta=(\textbf{r},\tau)$. The non-locality of the propagator is a consequence of the dimensional reduction performed in order to generate the (3+1)D propagator within (2+1)D space. A similar 
fact occurs when we integrate out
parts of the system degrees of freedom as, for instance, in  the Caldeira-Leggett model 
for dissipative quantum mechanics \cite{caldeiralegget}. 

Nevertheless, in spite of being non-local, the theories described by Eq.~(\ref{1}) do
respect causality. Indeed, it has been shown that the classic (retarded and advanced) Green functions
vanish outside of the light-cone for any $\alpha$, thus preserving causality \cite{marinorubenssup}. For the 
special case of $\alpha=1/2$, the classic Green functions reduce to a delta function on the light-cone surface \cite{marinorubenssup}. The interesting consequence of this property is that the theory will obey
Huygens principle in this case \cite{marinorubenssup,huygens}, while QED$_3$ does not obey it.

We see that the theories described by Eq.~(\ref{1}) satisfy causality despite the  apparent non-locality, but it is not a priori obvious whether they respect unitarity.
In the present work, we shall test the unitarity of those
theories through the application of the optical theorem.

\section{Unitarity at Tree Level}

Let us investigate here the unitarity of the theories given by Eq.~(\ref{1}) by considering the free Feynman propagator  (tree level)
in connection to the optical theorem.
We use the Feynman prescription $k^2 \rightarrow k^2 + i \varepsilon$
in order to define the gauge field propagator corresponding to (\ref{1})
\begin{equation}
G^{\mu\nu}_{F}(t, \textbf{r})  =\frac{1}{4} P^{\mu\nu} D_F(t, \textbf{r}) ,
\label{m0}
\end{equation}
where 
\begin{eqnarray}
P_{\mu\nu}=g_{\mu\nu}-\frac{\partial_{\mu}\partial_{\nu}}{\Box^2}
\end{eqnarray}
is the transverse projector, $g_{\mu\nu}$ is the Minkowski metric, and $ D_F(t, \textbf{r}) $ is the  corresponding scalar propagator, in the Minkowski space. Thus, we replace $\tau$ by $t$, therefore we have
\begin{equation}
D_F(t, \textbf{r})  = \int \frac{d \omega}{2\pi} \int \frac{d^2k}{(2\pi)^2} 
\frac{e^{-i\omega t}\ e^{i\textbf{k}\centerdot \textbf{r}}}{(\omega^2 - \textbf{k}^2+ i \varepsilon)^{1-\alpha}}.
\label{m1}
\end{equation}

This integral has been calculated in Ref.~\cite{marinorubenssup} (see Appendix 1 therein), yielding
\begin{equation}
D_F(t, \textbf{r})  =- \frac{C_\alpha}{(t^2-\textbf{r}^2- i \epsilon)^{1/2+\alpha}},
\label{m2}
\end{equation}
where 
\[
C_\alpha= \frac{2^{2\alpha-1/2}}{(2\pi)^{3/2}}\frac{\Gamma(\alpha+1/2)}{\Gamma(1-\alpha)}.
\]

In order to probe the unitarity of the theories described by Eq.~(\ref{1}),
let us first consider the scalar field. Later on we shall return to the vector field
case.

Taking the amplitude corresponding to the operator Eq.~(\ref{0}) evaluated between states
 $|i\rangle$ and $|f \rangle$, which is written as
$\langle i|T|f\rangle= (2\pi)^3 \delta^{3}(k_i-k_f) D_{if}$
and introducing a complete set of intermediate states $|x\rangle$ on the right-hand side (rhs), the above unitarity condition becomes  Ref.~\cite{matthew}
\begin{equation}
D^*_{if}-D_{if}= -i \sum_x  \int d\Phi \, (2\pi)^3 \delta^3(k_i-k_f)\, (D^*_{ix}D_{xf}),\label{optteo}
\end{equation}
where $d\Phi$ is the phase space factor, which is needed for dimensional reasons and also to ensure
that the sum over the intermediate states corresponds to the identity. The equation above is known as the generalized optical theorem.

Now, for $i \rightarrow f$, the amplitude $D_{ii}$ becomes the Feyman propagator, 
\[ D_{ii}=D_F
(t-t', \textbf{r}-\textbf{r}') 
\]
 which is given by Eq.~(\ref{m2}). Notice
that, in the Heisenberg picture $D_F(t-t', \textbf{r}-\textbf{r}') = \langle \textbf{r},t|\textbf{r}',t'\rangle$.

The unitarity condition, therefore, would lead to the equation 
\begin{eqnarray}
 D^*_F(t,\textbf{r})-  D_F(t,\textbf{r})&=&
\nonumber \\
-i   \int d\Phi \,  (2\pi)^3 \delta^3(0) \int \frac{d t_x}{2\pi} \int \frac{d^2r_x}{(2\pi)^2}
\nonumber \\
 D^*_F(t_x,\textbf{r}_x) D_F(t-t_x,\textbf{r}-\textbf{r}_x). 
 \label{unicond3}
\end{eqnarray}

Our strategy to test unitarity of a given theory will be to check whether the corresponding propagator
satisfies the optical theorem.
For this purpose, we Fourier transform the above equation to  energy-momentum space,
\begin{equation}
D^*_F(\omega,\textbf{k})-D_F(\omega,\textbf{k})=-i {\cal T}^{\gamma}D^*_F(\omega,\textbf{k})D_F(\omega,\textbf{k}),
\label{D}
\end{equation}
where $D_F(\omega,\textbf{k})$ is promptly obtained from Eq.~(\ref{m1}), it is given by
\begin{equation}
D_F(\omega,\textbf{k})=\frac{1}{(\omega^2-\textbf{k}^2+i\varepsilon)^{1-\alpha}}.  \label{genpropk}
\end{equation}

In the Eq.~(\ref{D}), we used the fact that
the phase space  integral combined with $\delta^3(0)$
yields $ {\cal T}^{\gamma}$,
where ${\cal T}$ is the characteristic time of the system and $\gamma=-2(1-\alpha)$ (see App.~A).

Defining  $ \chi_\alpha = (\omega^2-\textbf{k}^2+i\varepsilon)^{1-\alpha}$, we can write the equation above as
\begin{equation}
 \frac{2 \,{\rm Im}(\chi_\alpha)}{\chi^*_\alpha \chi_\alpha} =
 \frac{ {\cal T}^{-2(1-\alpha)}}{\chi^*_\alpha \chi_\alpha}.
\label{mm3}
\end{equation}
For unitarity to be respected, we must have
\begin{equation}
2 \,{\rm Im}( \chi_\alpha) = {\cal T}^{-2(1-\alpha)}.
\label{choice2}
\end{equation} 
However, since the rhs is a constant, for the above condition to be consistent, 
${\rm Im}(\chi_\alpha)$ must also be a
constant, in the limit $\varepsilon\rightarrow 0$. In other words, in that limit the left-hand side (lhs) can not be a function of $\lambda=\omega^2-\textbf{k}^2$ for Eq.~(\ref{choice2}) to be consistent.

In order to verify this condition, we introduce
a polar representation for $\chi_\alpha$, namely,
 $\chi_\alpha=(\rho \,e^{i\theta})^{1-\alpha}$, with $\rho^2(\lambda)=\lambda^2+\varepsilon^2$ and $\theta(\lambda)=\sin^{-1}(\varepsilon/\rho)$. Then, we require that 
\begin{equation}
\frac{d}{d\lambda}{\rm Im}(\chi_\alpha)=\frac{d }{d\lambda}\rho \sin[(1-\alpha)\theta]=0.
\end{equation} 

Calculating the derivative, we obtain 
\begin{equation}
\tan[\theta(\lambda)(1-\alpha)]=\tan[\theta(\lambda)],
\label{der}
\end{equation} 
which has an obvious solution $\alpha=0$. Indeed, it is clear
that for this value of $\alpha$, ${\rm Im}(\chi_{0}) = \varepsilon$ and therefore is independent of $\lambda$.

A less obvious solution is $\alpha=1/2$, which is valid because in this case Eq.~(\ref{der}) admits a solution
$\theta(\lambda) = 2\pi - \varepsilon$, which is compatible with the definition of $\theta(\lambda)$.
In this case, we also find ${\rm Im}(\chi_{1/2}) = \varepsilon$ (see App.~B).

We conclude that, for the theories whith $\alpha=0$ and $\alpha=1/2$,
the two sides of Eq.~(\ref{mm3}) would coincide consistently by identifying $2 \varepsilon$
with ${\cal T}^{-2}$. 
For other values of $\alpha$, ${\rm Im}(\chi_\alpha) $ would depend on $\lambda$ and, therefore, we 
would not be able to
 find a consistent solution of Eq.~(\ref{choice2})
satisfying the generalized optical theorem.

The demonstrations provided above were meant for the scalar theories associated with Eq.~(\ref{1}). The corresponding results
for the vector propagator Eq.~(\ref{m0}) then, follow straightforwardly by making ${\cal T}^{-2(1-\alpha)}/4\rightarrow {\cal T}'^{-2(1-\alpha)}$
and from the fact that the transverse projector has the property: $P^2=P$.

We conclude that out of the class of theories described by Eq. (\ref{1}), only the ones with $\alpha=0$ and $\alpha=1/2$, namely
QED$_3$ and PQED are unitary.

\section{Unitarity of PQED in the RPA approximation}

Next, we consider PQED, the case for which $\alpha=1/2$. As we have seen, it describes the 
EM interaction of the particles  coupled to it. Having graphene in mind 
we describe the electrons as massless Dirac fermions experiencing  the EM interaction
mediated by the gauge field $A_{\mu}$. The Lagrangian in this case reads \cite{gra} 
\begin{eqnarray}
{\cal L}&=& \frac{1}{4} F_{\mu \nu}\left[\frac{4}{\sqrt{-\Box}}\right] F^{\mu\nu} + \bar\psi\left(i\partial\!\!\!/+e\,\gamma^{\mu}\, A_{\mu}\right)\,\psi,
\end{eqnarray}
where $e$ is the dimensionless coupling constant, $\psi$ is the Dirac field, and $\gamma^{\mu}$ are Dirac matrices which can be either two or four dimensional, since we are in (2+1)D. 

The corrections to the gauge-field propagator are
expressed in terms of the
 the vacuum polarization $\Pi_{\mu\nu}(p)$. The one-loop expression
for this has been calculated in Ref. \cite{Luscher} and is given by
\begin{eqnarray}
\Pi_{\mu\nu}(k)=-\frac{e^2\sqrt{k^2}}{16}P_{\mu\nu}(k) +\frac{e^2}{2\pi}\left(n+ \frac{1}{2} \right)\epsilon_{\mu\nu\alpha}k^\alpha,
\label{eq1}
\end{eqnarray}
where $n$ is an integer. The result above is for two dimensional Dirac matrices.

According to Eq.~(\ref{m1}), the free gauge field propagator in momentum space reads
\begin{eqnarray}
G_{0,\mu\nu}(k)=\frac{1}{4} \frac{P_{\mu\nu}(k)}{\sqrt{k^2}}.
\end{eqnarray}

We include the vacuum polarization corrections by using the random phase approximation (RPA), where the corrected propagator
is given by the geometrical series
\begin{eqnarray}
G_{\mu\nu}=G_{0,\,\mu\alpha}\left[\delta_{\alpha,\nu}+\Pi^{\alpha\beta}G_{0,\,\beta\nu}+\right.\nonumber\\\left.
\Pi^{\alpha\beta}G_{0,\,\beta\sigma}\Pi^{\sigma\gamma}G_{0,\,\gamma\nu}+... \right ].
\end{eqnarray}

Because of the peculiar momentum dependence of the vacuum polarization tensor, the
corrected propagator has basically the same momentum dependence as the free one
\begin{eqnarray}
G_{\mu\nu}(k)=\frac{1}{\sqrt{k^2+i\epsilon}}\left(A_1 P_{\mu\nu}(k)+A_2 \frac{\epsilon_{\mu\nu\alpha}k^{\alpha}}{\sqrt{k^2}}\right), \label{fullprop}
\end{eqnarray}
where $A_1$ and $A_2$ are constants depending on the coefficients of the vacuum polarization tensor. Note that we use the Feynman prescription as we did before. Unitarity of the theory is guaranteed 
provided that the optical theorem Eq.~(\ref{optteo}) is still respected. 

The propagator above can be conveniently written as
\begin{eqnarray}
G_{\mu\nu}(k)= C_{\mu\nu}(k) D_F(k),
\end{eqnarray}
where 
\begin{equation}
C_{\mu\nu}(k)=A_1 P_{\mu\nu}(k)+A_2\frac{\epsilon_{\mu\nu\alpha}k^{\alpha}}{\sqrt{k^2}}, 
\label{oper}
\end{equation}
with $D_F(k)$ given by the Eq.~(\ref{m1}) for $\alpha=1/2$. 

The optical theorem now reads
\begin{eqnarray}
G^*_{\mu\nu}(t,\textbf{r})-G_{\mu\nu}(t,\textbf{r})&=&
\nonumber \\
-i   \int d\Phi \,  (2\pi)^3 \delta^3(0)  \int \frac{d t_x}{2\pi} \int \frac{d^2r_x}{(2\pi)^2}
\nonumber \\
G^*_{\mu\alpha}(t_x,\textbf{r}_x) G_{\alpha\nu}(t-t_x,\textbf{r}-\textbf{r}_x).  
\label{optteo22}
\end{eqnarray}

Next, we adopt the same strategy as for the non-interacting case and perform a Fourier transform in both sides of the above equation, again, considering that the Fourier transform of a convolution is a product. We obtain
\begin{equation}
G^*_{\mu\nu}(\omega,\textbf{k})-G_{\mu\nu}(\omega,\textbf{k})= 
-i {\cal T}^{-1}G^*_{\mu\alpha}(\omega,\textbf{k})G_{\alpha\nu}(\omega,\textbf{k}).
\label{optteo33}
\end{equation}
 The lhs of Eq.~(\ref{optteo33}) is  given by
\begin{eqnarray}
\frac{C_{\mu\nu}(k)2i  \,{\rm Im}(\chi_{1/2})}{[(\omega^2-\textbf{k}^2)^2+\epsilon^2]^{1/2}},
\label{lhsoptteo2}
\end{eqnarray}
whereas the rhs of Eq.~(\ref{optteo33}) reads
\begin{eqnarray}
\frac{-i{\cal T}^{-1} C_{\mu\alpha}(k)C_{\alpha\nu}(k)}{[(\omega^2-\textbf{k}^2)^2+\epsilon^2]^{1/2}}, 
\label{rhsoptteo2}
\end{eqnarray}
where
\begin{eqnarray}
C_{\mu\nu}^2(k)= (A_1^2-A_2^2) P_{\mu\nu}(k)-2 A_1 A_2 \frac{\epsilon_{\mu\nu\alpha}k^{\alpha}}{\sqrt{k^2}} \label{oper2}.
\end{eqnarray}

We now consider Eq.~(\ref{lhsoptteo2}) and Eq.~(\ref{rhsoptteo2}). Since both are proportional to the operators $P_{\mu\nu}(k)$ and $\epsilon_{\mu\nu\alpha}k^{\alpha}/\sqrt{k^2}$, therefore, we have to compare the corresponding coefficients of both terms. 
Using the result of App.~B, we conclude that the optical theorem  will be obeyed and consequently, unitarity preserved, provided we make the choices
\begin{equation}
(2\ \varepsilon)^{1/2} = \frac{A_1^2-A_2^2}{2A_1} {\cal T}^{-1},
\end{equation}
in the $P_{\mu\nu}(k)$ term and
\begin{equation}
(2\ \varepsilon')^{1/2} = A_1  {\cal T}^{-1},
\end{equation}
in the $\epsilon_{\mu\nu\alpha}k^{\alpha}/\sqrt{k^2}$ term.

This concludes our proof of the unitarity of PQED of massless electrons in the RPA approximation.

\section{Beyond the RPA approximation}

Within the RPA approximation, the one-loop expression for the vacuum polarization tensor,
Eq.~(\ref{eq1}) is used in the geometrical series that corrects the free propagator of the gauge field.
This approach can be improved  by adding the two-loop correction for the vacuum polarization tensor, as calculated by Teber \cite{Teber}, 
\begin{equation}
\Pi_{\mu\nu}^{(2)}(k)=-\frac{\sqrt{k^2}}{16}\,\left(\frac{92-9\pi^2}{18\pi}\right)\,\alpha_g\,P_{\mu\nu}\,, 
\label{pi2}
\end{equation}
where  $\alpha_g\approx 300/137 = 2.189$ is the fine structure constant of graphene.
Considering that $(92-9\pi^2)/18\pi\approx 0.056$, we see that the two-loop correction is sensible. There is no correction to the Chern-Simon term due to the Coleman-Hill theorem \cite{ch}.

Observe that, remarkably, the two-loop correction has precisely the same functional dependence as
the one-loop one. As a consequence, the only effect of the two-loop correction to the
vacuum polarization is to redefine the constant $A_1$ in Eq.~(\ref{fullprop}). Therefore, it immediately follows
that the optical theorem, and consequently, unitarity are respected in the two-loop extension of the RPA
approximation.

\section{Conclusions}

We have tested the unitarity of a class of field theories in 2+1D containing fractional powers ($1-\alpha$) of the
d'Alembertian operator, which despite being nonlocal, respect causality. QED$_3$ and PQED are particular
cases, respectively, with $\alpha =0$ and $\alpha=1/2$. 

Our strategy is to verify whether the propagator
satisfies the optical theorem. We first considered the free propagator for generic $\alpha$ and showed that
only for $\alpha = 0$ and $\alpha=1/2$, namely, for  QED$_3$ and PQED, unitarity is respected. Inspection of the propagators in Eqs.~(\ref{m2}) and (\ref{genpropk}) shows that one theory is dual to the other. Indeed, for $\alpha=0$ the exponent in $D_F(\omega,\textbf{k})$ is unity, whereas the one in $D_F(t,\textbf{r})$ is $1/2$. For $\alpha=1/2$, the same occurs, but with $\textbf{k}\rightarrow \textbf{r}$ and $\omega\rightarrow t$.

We then considered the case of PQED coupled to massless Dirac fermions, which is the model for graphene.
We have shown that the propagator corrected both within the RPA approximation and in its two-loop
extension satisfy the optical theorem, hence unitarity is preserved in both cases.  

\section{Acknowledgments}

This work was supported in part by CNPq (Brazil), CAPES (Brazil), FAPERJ (Brazil), The Netherlands Organization for Scientific Research (NWO) and by the Brazilian
government project Science Without Borders.
We are grateful to G.'t Hooft for interesting discussions.

\section{Appendix A: The Phase Space Factor}

Here we are going to determine the phase space factor \cite{peskin}. Let us consider Eq.~(\ref{unicond3})
and write
\begin{equation}
\int d\Phi \,  (2\pi)^3 \delta^3(0) \equiv {\cal T}^\gamma,
\end{equation}
where ${\cal T}$ is the characteristic time scale of the system.
For dimensional reasons, we have $\gamma +3 = 2(\alpha + 1/2)$ and consequently $\gamma = -2(1-\alpha)$.
This justifies the $\gamma$-dependence in Eq.~(\ref{D}).

\section{Appendix B: The  ${\rm Im} (\chi_\alpha) $}

Let us show here that, for $\alpha= 0, 1/2$, indeed, the expression of  ${\rm Im} (\chi_\alpha) $ relevant
for the optical theorem, is given by $\varepsilon,\varepsilon^{1/2} $, 
respectively, and therefore just depends on $\varepsilon$.

Using $ \chi_\alpha\equiv(\omega^2-\textbf{k}^2+i\varepsilon)^{1-\alpha}$, we have,
for $\alpha = 0$,  $ \chi_0 = (\omega^2-\textbf{k}^2+i\varepsilon)$ and evidently  $ {\rm Im} (\chi_0) = \varepsilon \propto {\cal T}^{-2}$.

For the case $\alpha = 1/2$, notice that the condition for the optical theorem to be satisfied is
\begin{equation}
 \frac{2 \,{\rm Im}(\chi_{1/2})}{[(\omega^2-\textbf{k}^2)^2+\varepsilon^2]^{1/2}} =
 \frac{K {\cal T}^{-1}}{[(\omega^2-\textbf{k}^2)^2+\varepsilon^2]^{1/2}},
\label{B1}
\end{equation}
for some dimensionless constant K.
Squaring this equation and multiplying both the numerators by $\varepsilon$, we obtain both sides proportional
to $\delta(\omega^2-\textbf{k}^2)$. As a consequence, we must equate the numerators at $\omega^2-\textbf{k}^2=0$,
namely, 
\begin{equation}
2\ {\rm Im}(\chi_{1/2})\Big |_{\omega^2=\textbf{k}^2} =(2 \ \varepsilon)^{1/2} =
K {\cal T}^{-1},
\label{B2}
\end{equation}
which completes the proof for $\alpha = 1/2$.

\end{document}